\documentclass[aps,prl,showpacs,amsmath,amssymb,amsfonts,twocolumn,superscriptaddress,checklength]{revtex4-2}

\usepackage{graphicx}% Include figure files
\usepackage{dcolumn}% Align table columns on decimal point
\usepackage{bm}% bold math
\usepackage[colorlinks=true,linkcolor=blue,citecolor=blue]{hyperref}
\usepackage{xcolor}
\usepackage{ulem}

\begin{document}

\author{Zi-Qi Wang}
\email{These authors contributed equally to this work.}
\affiliation{Zhejiang Key Laboratory of Micro-Nano Quantum Chips and Quantum Control, State Key Laboratory for Extreme Photonics and Instrumentation, and School of Physics, Zhejiang University, Hangzhou 310027, China}
\author{Yi-Ming Sun}
\email{These authors contributed equally to this work.}
\affiliation{Zhejiang Key Laboratory of Micro-Nano Quantum Chips and Quantum Control, State Key Laboratory for Extreme Photonics and Instrumentation, and School of Physics, Zhejiang University, Hangzhou 310027, China}
\author{Yao-Dong Hu}
\email{These authors contributed equally to this work.}
\affiliation{Zhejiang Key Laboratory of Micro-Nano Quantum Chips and Quantum Control, State Key Laboratory for Extreme Photonics and Instrumentation, and School of Physics, Zhejiang University, Hangzhou 310027, China}
\author{Yi-Pu Wang}
\email{yipuwang@zju.edu.cn}
\affiliation{Zhejiang Key Laboratory of Micro-Nano Quantum Chips and Quantum Control, State Key Laboratory for Extreme Photonics and Instrumentation, and School of Physics, Zhejiang University, Hangzhou 310027, China}
\author{Rui-Chang Shen}
\affiliation{Zhejiang Key Laboratory of Micro-Nano Quantum Chips and Quantum Control, State Key Laboratory for Extreme Photonics and Instrumentation, and School of Physics, Zhejiang University, Hangzhou 310027, China}
\affiliation{Department of Physics, The Chinese University of Hong Kong, Shatin, Hong Kong SAR, China}
\author{Wei-Jiang Wu}
\affiliation{Zhejiang Key Laboratory of Micro-Nano Quantum Chips and Quantum Control, State Key Laboratory for Extreme Photonics and Instrumentation, and School of Physics, Zhejiang University, Hangzhou 310027, China}
\affiliation{Department of Physics, The Chinese University of Hong Kong, Shatin, Hong Kong SAR, China}
\author{J. Q. You}
\email{jqyou@zju.edu.cn}
\affiliation{Zhejiang Key Laboratory of Micro-Nano Quantum Chips and Quantum Control, State Key Laboratory for Extreme Photonics and Instrumentation, and School of Physics, Zhejiang University, Hangzhou 310027, China}

\title{Enhancement of signal-to-noise ratio at a high-order exceptional point of coherent perfect absorption}

	\begin{abstract}
\noindent\textbf{Abstract:}Exceptional points (EPs) in non-Hermitian systems offer a remarkably strong response to weak perturbations, but the nonorthogonal nature of the corresponding eigenvectors causes noise to diverge, hindering EPs practical application. Here, we report a twelve-fold enhancement of signal-to-noise ratio (SNR) in magnetic field sensing enabled by a third-order EP of coherent perfect absorption (CPA EP3) in a passive cavity magnonic system. This non-Hermitian magnonic platform comprises two identical yttrium iron garnet (YIG) spheres coherently coupled to a cavity mode, in which the CPA EP3 is realized by engineering the three-mode loss to form a pseudo-Hermitian absorption Hamiltonian. By independently tailoring the absorption EP apart from the resonance EP, the system circumvents the noise divergence caused by eigenbasis collapse. Notably, we harness the sensitivity of the minimum output intensity near CPA to perturbations, yielding a seventy-fold SNR improvement and a 400-fold increase in responsivity compared with non-CPA system. A comprehensive noise analysis over one hundred repeated measurements confirms the suppression of frequency noise near the CPA EP3. This demonstrates that our scheme not only avoids the noise divergence plaguing conventional higher-order EP sensors but also provides a general strategy to exploit both CPA and EP for SNR enhancement in passive non-Hermitian systems.

	\end{abstract}

	\maketitle
\vspace{1em}
\noindent\textbf{\large Introduction}\\
An exceptional point (EP) is a singularity in non-Hermitian systems, where both eigenvalues and eigenvectors coalesce~\cite{Bender2007,Rotter2009,Shen2018,Leykam2017,Bergholtz2021,Chrisrian2010,Lin2011,Peng2014,Zhen2015,Guo2009,Doppler2016,Feng2014,Hodaei2014,Brandstetter2014,Lin2016,Dembowski2001,Xu2016,Hassan2017,Zhou2018,Ozdemir2019,Miri2019,Parto2020,Jing2014,Ding2016,Liu2019,Tang2020,Kawabata2019,Rao2024,Lambert2025}. In contrast to the linear response in Hermitian degeneracies, the eigenfrequency splitting follows a distinct $\epsilon^{1/n}$ scaling with the perturbation strength $\epsilon$ near an $n$th-order EP (EP$n$). This drastically enhanced splitting has garnered great interest in sensor schemes ~\cite{Wiersig2014,Hodaei2017,Chen2017,Zhong2019,Lai2019,Wiersig2020,Budich2020} since it can considerably improve system response to weak perturbations without sacrificing the operational range. While pronounced splitting has been experimentally verified at second-~\cite{Chen2017} and higher-order EPs~\cite{Hodaei2017,xiao_enhanced_2019}, recent research findings have revealed that the EP sensor fails to deliver signal-to-noise ratio (SNR) improvements in practical implementations, due to the diverging noise steming from eigenvector nonorthogonality~\cite{Wiersig2020,loughlin_exceptional-point_2024,lau_fundamental_2018,xia_2025,wang_petermann-factor_2020,langbein_no_2018,duggan_limitations_2022,kononchuk_exceptional-point-based_2022,wiersig_prospects_2020,Wang_2023}. 
 
To advance the performance of EP-based sensors, considerable efforts have been devoted to overcoming the SNR limitations~\cite{kononchuk_exceptional-point-based_2022,bai_nonlinear_2023,bai_observation_2024,peters2022exceptional,lau_fundamental_2018,Kottos_2023,xia_2025,zhao_2024}, such as introducing nonreciprocal coupling~\cite{lau_fundamental_2018,zhao_2024} and nonlinearity~\cite{bai_nonlinear_2023,bai_observation_2024,peters2022exceptional} into the system. But these approaches inevitably engender the instabilities and remain controversial \cite{zheng_noise_2025}. A particularly promising approach is to exploit singularities arising in non-Hermitian scattering and in open systems driven far from thermal equilibrium~\cite{Sergeev_2021}. As the system's response to perturbations in practical measurements is manifested in the observed spectra, the signal degeneracies can be engineered to circumvent the resonance EP. This scheme has only been realized in the EP2 of an active electromechanical accelerometer system~\cite{kononchuk_exceptional-point-based_2022}, where a threefold SNR enhancement was achieved. Recently, a fully passive non-Hermitian magneto-optical EP2 sensor has also been shown to enhance the SNR by exploiting loss-induced EP response, thereby avoiding the noise penalty associated with gain-based schemes~\cite{xia_2025}.

\begin{figure*}
\centering
\includegraphics[width=1\textwidth]{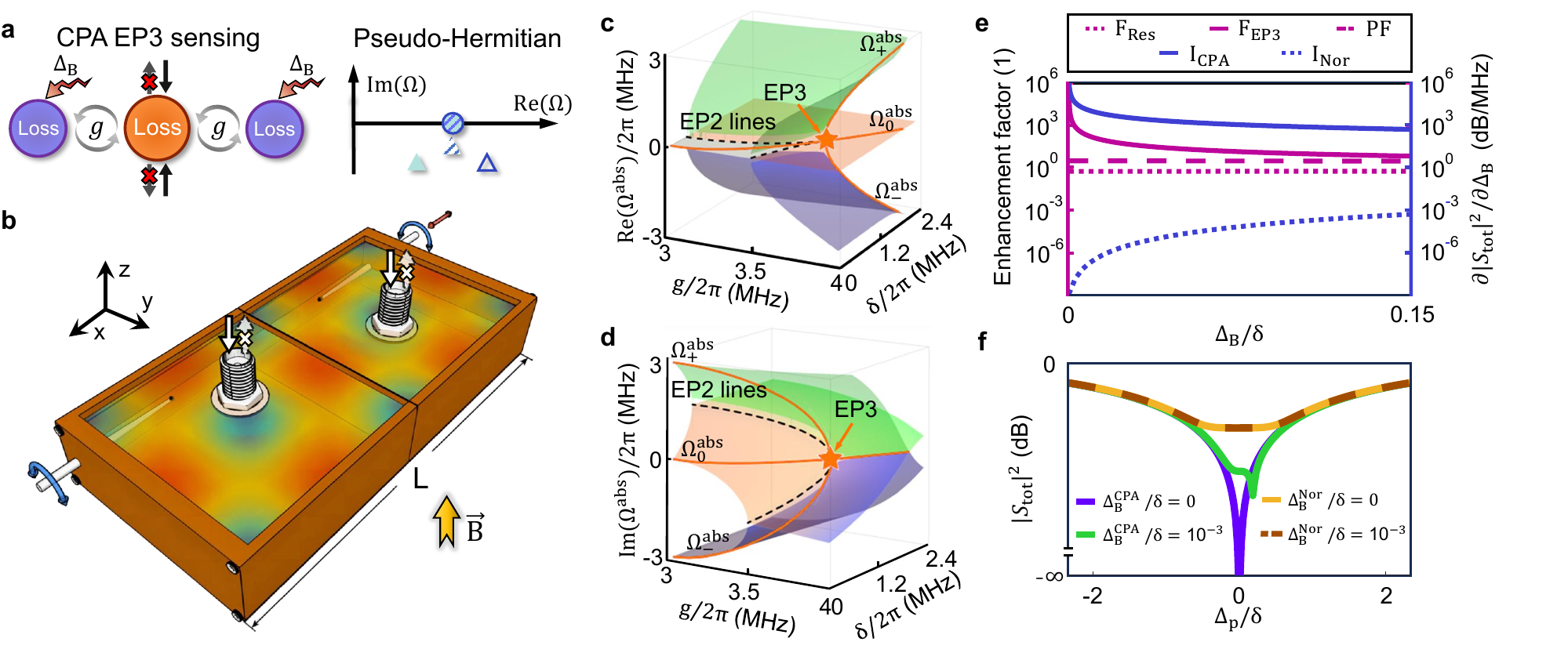}
\caption{\textbf{Sensing via higher-order exceptional point of coherent perfect absorption in a cavity magnonic system.} 
\textbf{a}, Left: schematic of a third-order exceptional point of coherent perfect absorption (CPA EP3) sensing, where two magnon modes (purple), with equal dissipation rates and opposite detunings $\pm\delta$, are coherently coupled to a cavity mode (orange). Two coherent probe signals are injected into the cavity, while frequency perturbations $\Delta_B$ are applied to the magnon modes. Right: Eigenvalues of $H_{\rm abs}$ (circles) and $H_{\rm res}$ (triangles) in the complex plane, where pseudo-Hermiticity enables the coalescence of three CPA solutions.  
\textbf{b}, Two 0.5~mm-diameter YIG spheres are placed in a 3D cavity, where the step motors allow both displacement and rotation. The magnetic-field distribution of the cavity mode $\text{TE}_{102}$ is shown as a color map.  
\textbf{c,d}, Real and imaginary parts of the eigenvalues of $H_{\rm abs}$ versus the coupling strength $g/2\pi$ and frequency detuning $\delta/2\pi$, illustrated by the green, orange, and purple surfaces, respectively. The black dashed curves indicate the EP2 lines, and their intersection point corresponds to the EP3 (red star).  
\textbf{e}, Eigenfrequency response of $\Omega^{\rm res}$ ($F_{\rm res}$, pink dotted) and $\Omega^{\rm abs}$ ($F_{\rm EP3}$, pink solid), Petermann factor (PF, pink dashed), together with the minimum output intensity response at CPA EP3 ($I_{\rm CPA}$, blue solid) and under normal conditions ($I_{\rm nor}$, blue dashed) versus perturbation $\Delta_B$.  
\textbf{f}, Numerically calculated output spectra versus $\Delta_p=\omega-\omega_c$ for $\Delta_B/\delta=0$ and $10^{-3}$, comparing the system operating at the CPA EP3 (purple and green lines) with the normal configuration (orange and brown dashed lines).
}
	\label{fig1}
			 
\end{figure*}
To further considerably improve SNR enhancement, we experimentally realize a coherent perfect absorption (CPA) EP3 in a purely passive cavity–magnonic system, achieving a {\it twelve-fold} enhancement of the SNR for frequency-based magnetic-field sensing. The CPA degeneracy offers a readily identifiable spectral feature that facilitates precise frequency extraction. The underlying noise resilience mechanism relies on intentionally offsetting this absorption EP from resonance EP of the system. Since the valley frequency in the absorption spectrum is determined by the absorption Hamiltonian, its nonlinear response at the absorption EP can be used to boost sensing while avoiding extra noise from eigenvector nonorthogonality. To realize CPA EP3, we design a pseudo-Hermitian absorption Hamiltonian~\cite{Mostafazadeh2002,Mostafazadeh2002-1,Rivero2020,Zhang2019} and observe a fifteen-fold increase in responsivity to small perturbations using two properly arranged yttrium iron garnet (YIG) spheres~\cite{Zhang2017,ziqi2022,Han2023}. The noise characteristics of the sensor are evaluated through 100 repeated measurements, with the noise quantified by the one-standard deviation. Importantly, the frequency noise near the CPA EP3 remains stable without any abnormal increase.

However, progressive enhancement of the SNR through increasing EP orders is fraught with formidable difficulty. In this work, we develop a CPA-based sensing mechanism by exploiting the giant variation of the minimum output intensity in the vicinity of the CPA~\cite{Chong2010,PhysRevLett.106.093902,Sun2014,Baranov2017,Soleymani2022,Zhang2023}. Its core principle is aptly visualized through the change in the product of the distances between absorption Hamiltonian’s eigenvalues on the complex plane and the corresponding frequencies on the real axis. We find a 400-fold enhancement in its response to perturbations at the CPA compared to the non-CPA case. With the spectral output noise substantially reduced near the CPA, the SNR of the minimum output intensity is enhanced seventy-fold. Our work not only effectively avoids the noise divergence in higher-order EP sensing, but also opens an alternative route to greatly enhance the SNR by harnessing both CPA and EP in a purely passive non-Hermitian system.

\begin{figure*}[htbp]
	\centering
	\includegraphics[width=0.999\textwidth]{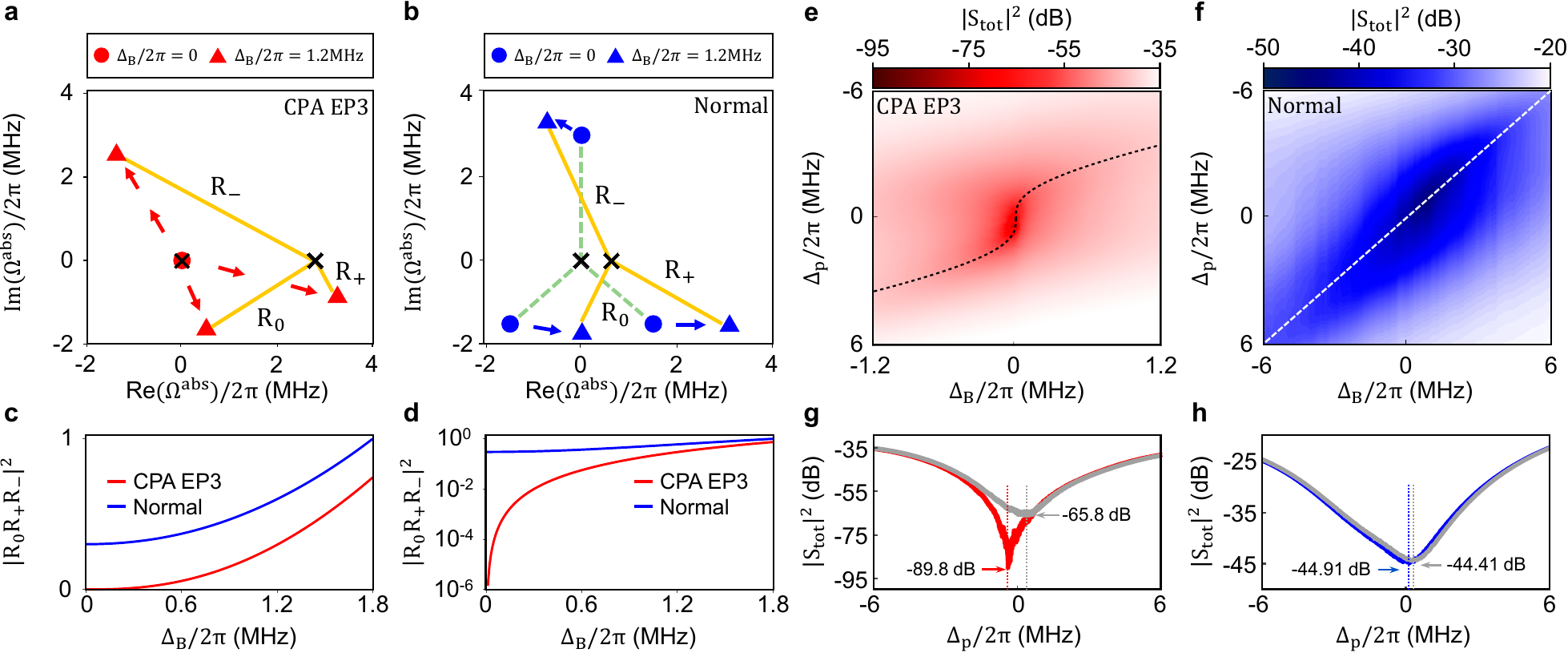}
	\caption{\textbf{Output spectra of the system under the third-order exceptional point of coherent perfect absorption and normal conditions.} 
\textbf{a,b}, Eigenvalues $\Omega^{\rm abs}$ on the complex plane under third-order exceptional point of coherent perfect absorption (CPA EP3) (\textbf{a}) and normal (\textbf{b}) conditions, with perturbations $\Delta_B/2\pi$= 0 (circles) and 1.2 MHz (triangles). The cross on the x-axis marks $\omega_{\rm min}$ at each $\Delta_B$. The orange solid and green dashed lines represent the distances $R_{0,\pm}$ from $\Omega_{0,\pm}^{\rm abs}$ to the corresponding $\omega_{\rm min}$ at $\Delta_B/2\pi$= 0 and 1.2 MHz, respectively. \textbf{c,d}, Numerical calculations of $|R_0R_+R_-|^2$ versus $\Delta_B$ under CPA EP3 and normal conditions, shown in normalized linear scale (\textbf{c}) and logarithmic scale (\textbf{d}), respectively.\textbf{e,f}, Total output spectra versus the perturbation $\Delta_B$ under the CPA EP3 (\textbf{e}, $g=g_{\rm EP3}=2\pi \times 3.46$~MHz) and normal conditions (\textbf{f}, $g=0.9g_{\rm EP3}$). The dashed lines indicates the frequency corresponding to the minimum output intensity. \textbf{g,h}, Total output spectra in \textbf{g} and \textbf{h} are extracted from \textbf{e} and \textbf{f} at $\Delta_B/2\pi=0$ and $0.06$~MHz, respectively.}\label{fig:2}
\end{figure*}

\vspace{1em}
\noindent\textbf{\large Results}\\
\noindent\textbf{System and model}\\
Our CPA EP3 sensing model is illustrated in Fig.~\ref{fig1}\textbf{a}, where a lossy mode coupled to two probing channels interacts with two modes of identical dissipation $\gamma$ via equal strength $g$. Experimentally, we implement this scheme with a three-dimensional rectangular cavity ($50\times25\times7.5~\rm{mm}^3$) hosting two 0.5~mm-diameter YIG spheres (Fig.~\ref{fig1}\textbf{b}). The cavity ports are connected to a vector network analyzer (VNA) for both excitation and measurement, with the incident signals’ power ratio and phase difference precisely tuned by a variable attenuator and a phase shifter. Each YIG sphere is mounted on a plastic rod for position and crystal-axis adjustment. Under a bias magnetic field $\textbf{\textit{B}}$ applied along the $z$-direction, the spheres support magnon modes. We focus on the Kittel mode, namely the uniform spin precession. When the magnon modes are detuned from the cavity frequency $\omega_{c}$, the resonance Hamiltonian $H_{\rm res}$ of the cavity–magnon system without probe signals takes the form
\begin{eqnarray}
	H_{\rm{res}}/\hbar=
	\left(
	\begin{matrix}
		-i\kappa_c & g & g\\
		g & \delta_1-i\gamma & 0\\
		g & 0 & \delta_2-i\gamma\\
	\end{matrix}\right),
	\label{eq:1}
\end{eqnarray}
where $\delta_{1(2)} = \omega_{1(2)}-\omega_{c}$ denotes the frequency detuning between each magnon mode and the cavity mode, and $\kappa_{c}\equiv \kappa_1+\kappa_2+\kappa_{\rm int}>0$ is the total cavity dissipation rate, with $\kappa_{1,2}$ being the external damping rates arising from the coupling between the cavity and the ports, and $\kappa_{\rm int}$ being the intrinsic damping rate. When two in-phase input signals with frequency $\omega$ are injected into the cavity, the output intensity is expressed as (see Supplementary Materials)
\begin{equation}\label{stot}
	\left| {\rm S_{\rm tot}(\omega)} \right|^{2} =(p+1) \frac{|{\bf det}(\omega\hat{I}-H_{\rm abs})|^2}{|{\bf det}(\omega\hat{I}-H_{\rm res})|^2},
\end{equation}
where $p$ is the power ratio of two input signals, $H_{\rm{abs}}=H_{\rm{res}}+iD^{\dagger}D$ denotes the absorption Hamitonian and $D= \begin{pmatrix}
	\sqrt{2\kappa_{1}} & 0 & 0 \\
	\sqrt{2\kappa_{2}} & 0 & 0 \\
\end{pmatrix}$ characterizes the coupling between the cavity mode and two ports. The absorption Hamiltonian $H_{\rm{abs}}$ takes a form similar to $H_{\rm{res}}$, except that the cavity dissipation is given by $\kappa_{c}^{\rm abs}=-\kappa_{1}-\kappa_{2}+\kappa_{\rm int}$, which can be compensated through the interaction with the two coupling channels. As shown in Eq.~(\ref{stot}), the absorption solutions $\Omega^{\rm abs}$ are governed by the eigenvalues of $H_{\rm{abs}}$. To realize their degeneracy, we employ two identical YIG spheres with opposite detunings, $\delta_1=-\delta_2\equiv\delta$, relative to the cavity mode. By tuning $\kappa_{1,2}$ via the insertion depth of the port antenna pins, the cavity mode can effectively act as a gain mode with $\kappa_{c}^{\rm abs}=-2\gamma$ (see Supplementary Materials). Under this condition, $H_{\rm{abs}}$ reduces to
\begin{eqnarray}\label{eq:3}
H_{\rm{abs}}/\hbar =
\left(
\begin{matrix}
	2i\gamma & g & g\\
	g & \delta-i\gamma & 0\\
	g & 0 & -\delta-i\gamma\\
\end{matrix}\right).
\end{eqnarray}

For the experimental realization of CPA, corresponding to a vanishing output amplitude, the eigenvalues of $H_{\rm{abs}}$ must be purely real (${\rm Im}(\Omega^{\rm abs})=0$). This condition is satisfied by setting $\delta=\sqrt{g^2-\gamma^2}$, under which $H_{\rm{abs}}$ becomes {\it pseudo-Hermitian}~\cite{Mostafazadeh2002,Mostafazadeh2002-1,Kawabata2019,Rivero2020} with eigenvalues $\Omega_0^{\rm abs}=0$ and $\Omega_{\pm}^{\rm abs}=\pm\sqrt{3g^2-4\gamma^2}$. To examine their coalescence, we plot ${\rm Re}(\Omega_{0,\pm}^{\rm abs})$ and ${\rm Im}(\Omega_{0,\pm}^{\rm abs})$ as functions of $g$ and $\delta$ in Figs.~\ref{fig1}\textbf{c} and \ref{fig1}\textbf{d}. The EP3 (red star) appears at the intersection of two EP2 curves (black dashed lines), where all three eigenvalues coalesce. From Figs.~\ref{fig1}\textbf{c} and \ref{fig1}\textbf{d}, all eigenvalues remain real for $g>g_{\rm EP3}=2\gamma/\sqrt{3}$, resembling the behavior in parity-time (PT) symmetric systems~\cite{Zhang2019}.

\begin{figure*}[t]
	\centering
	\includegraphics[width=0.996\textwidth]{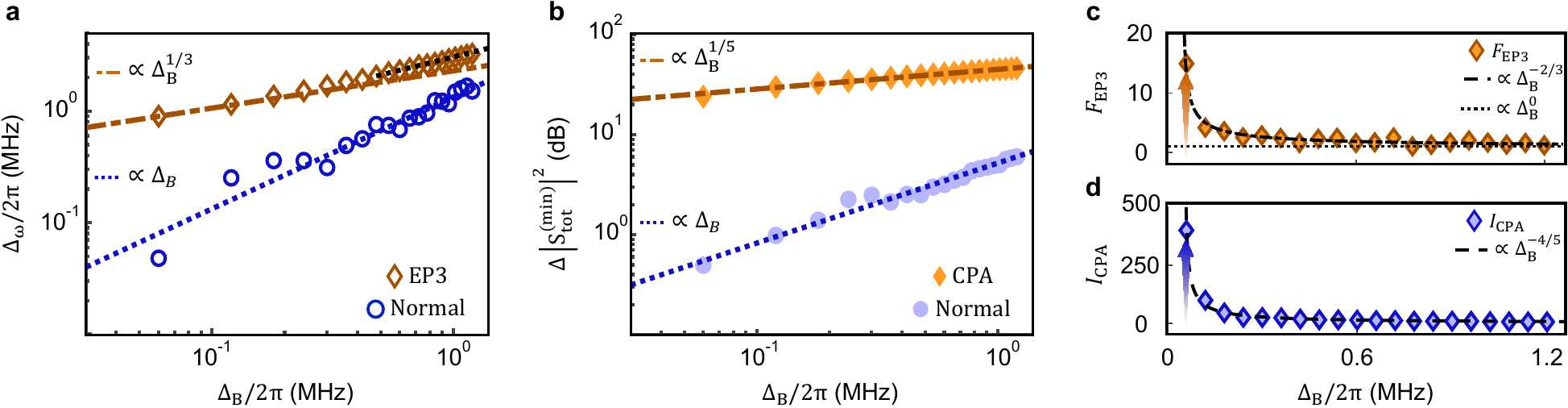}
	\caption{\textbf{Measured responses in frequency and minimum output intensity to magnetic-field variations.} \textbf{a,b}, Log-log plots of the changes in frequency $\Delta_\omega/2\pi$ (\textbf{a}) and intensity $\Delta|S_{\rm tot}^{({\rm min})}|^2$ (\textbf{b}) of the minimum output spectra versus the perturbation $\Delta_B/2\pi$ under the third-order exceptional point of coherent perfect absorption (CPA EP3) and normal conditions. The slopes 1/3, 1/5, and 1 are obtained by numerical fitting with the system parameters. \textbf{c,d}, Response enhancement factors at the CPA EP3 for frequency (\textbf{c}) and minimum output intensity (\textbf{d}).}
	\label{fig:3}
\end{figure*}

Interestingly, this higher-order singularity manifests exclusively in $H_{\rm abs}$, whereas the eigenvalues $\Omega^{\rm res}$ of the resonance Hamiltonian $H_{\rm res}$ remain non-degenerate at the CPA EP3, as shown in the right panel of Fig.~\ref{fig1}\textbf{a}. Under a perturbation $\Delta_B$ applied to the two magnon modes, the resonance response $F_{\rm res}\equiv \partial\Delta\Omega^{\rm res}/\partial\Delta_B$ remains constant with $\Delta_B$ (pink dotted line, Fig.~\ref{fig1}\textbf{e}), where $\Delta\Omega^{\rm res}$ denotes the frequency shift. In practice, however, sensing is governed by the frequency shift of the absorption minimum, determined by $\Omega^{\rm abs}$ in Eq.~(\ref{stot}). At the CPA EP3, this valley frequency shift $F_{\rm EP3}$ follows a cubic-root dependence on $\Delta_B$ (pink solid line, Fig.~\ref{fig1}\textbf{e}). Note that the apparent enhancement of the divergent $F_{\rm EP3}$ near $\Delta_B = 0$, as shown in Fig.~\ref{fig1}\textbf{f}, is limited by the minimum step size adopted in the numerical differentiation. To provide a more intuitive demonstration of the underlying nonlinear response, we directly calculate the output spectra as a function of $\Delta_p = \omega - \omega_c$. In this case, the system operating at the CPA EP3 (purple and green curves) exhibits a markedly larger frequency shift compared with the normal configuration (orange and brown curves). Crucially, the system’s eigenbasis neither collapses nor exhibits increasing overlap at the CPA EP3, as quantified by the Petermann factor (PF; pink dashed line in Fig.~\ref{fig1}\textbf{e}), in sharp contrast to conventional EPs, where a divergent PF gives rise to excess noise. Owing to the nonlinear response of $\Omega^{\rm abs}$ and the absence of eigenbasis collapse, CPA-EP3 sensing achieves a substantially improved SNR over conventional EP sensors.

Moreover, the minimum output intensity in Fig.~\ref{fig1}\textbf{f} also exhibits a pronounced change with perturbations in the CPA regime. To characterize this behavior, we define the response of the minimum output intensity as
	$I \equiv \partial \Delta |S_{\rm tot}^{({\rm min})}|^2 / \partial \Delta_B$,
	with $\Delta |S_{\rm tot}^{({\rm min})}|^2 \equiv
	10\log_{10}\!\left|{S'}_{\rm tot}^{({\rm min})}(\Omega^{\rm abs'})\right|^2
	- 10\log_{10}\!\left|S_{\rm tot}^{({\rm min})}(\Omega^{\rm abs})\right|^2$,
	where $\left|{S'}_{\rm tot}^{({\rm min})}(\Omega^{\rm abs'})\right|^2$ and
	$\left|S_{\rm tot}^{({\rm min})}(\Omega^{\rm abs})\right|^2$ denote the minimum
	output intensities with and without the perturbation $\Delta_B$, respectively.
	As shown in Fig.~\ref{fig1}\textbf{e}, we evaluate this response under CPA
	($I_{\rm CPA}$, blue solid line) and non-CPA ($I_{\rm Nor}$, blue dashed line)
	conditions. The $I_{\rm CPA}$ closely resembles the behavior at an EP. For sufficiently small perturbations, the variation of $I_{\rm CPA}$ is much larger than $I_{\rm Nor}$, and within the same perturbation range it even exceeds the resonance frequency shift at the EP3. To clarify the underlying mechanism of this scheme, we plot the eigenvalues $\Omega_{0,\pm}^{\rm abs}$ versus $\Delta_B$ in both CPA EP3 and normal cases on the complex plane, as shown in Figs.~\ref{fig:2}\textbf{a} and \ref{fig:2}\textbf{b}. From Eq.~(\ref{stot}), it is clear that the numerator governs the overall variation of $\left| {\rm S_{\rm tot}(\omega)} \right|^{2}$ for small $\Delta_B$, which can be recast as $|R_0R_+R_-|^2$ with $|R_{0,\pm}|^2\equiv|(\omega_{\rm min}-\Omega_{0,\pm}^{\rm abs})|^2$. The $\omega_{\rm min}$ corresponds to the real frequency of the minimum output intensity, which is numerically aquired and indicated as the cross in  Figs.~\ref{fig:2}\textbf{a} and \ref{fig:2}\textbf{b}. $R_{0,\pm}$ denotes the distance from $(\text{Re}(\Omega_{0,\pm}^{\rm abs}),\text{Im}(\Omega_{0,\pm}^{\rm abs}))$ to $(\omega_{\rm min}, 0)$. In this framework, the variation of the minimum output intensity can be directly visualized by the evolution of the distance product $|R_0R_+R_-|$. In the CPA EP3 case, the $\omega_{\rm min}$ coincides with the degenerate $\Omega_{0,\pm}^{\rm abs}$ at $\Delta_B = 0$, leading to a vanishing $|R_0R_+R_-|$ and thereby zero output intensity (red circle in Fig.~\ref{fig:2}\textbf{a}). Owing to the unique combination of CPA and EP3, a slight $\Delta_B$ substantially shifts the three eigenvalues and $\omega_{\rm min}$ in the complex plane, facilitating a dramatic change in $R_{0,\pm}$ (from {\it zero} to {\it nonzero} value) and their product (orange solid lines in Fig.~\ref{fig:2}\textbf{a}). While in the non-CPA EP3 case,  the initially nonzero $|R_{0,\pm}|$ for the separated $\Omega_{0,\pm}^{\rm abs}$ slightly change under perturbation, as indicated by the green dashed and orange solid lines in Fig.~\ref{fig:2}\textbf{b}. We plot the $|R_0R_+R_-|^2$ versus ${\Delta_B}$ under both CPA and non-CPA conditions on normalized linear (Fig.~\ref{fig:2}\textbf{c}) and logarithmic (Fig.~\ref{fig:2}\textbf{d}) scale, respectively. Although the CPA’s advantage is not apparently revealed on the linear scale due to the limited absolute change in $|R_0R_+R_-|^2$, it becomes conspicuous on a logarithmic scale, as shown by the decibel spectrum around the CPA (Fig.~\ref{fig1}\textbf{f}). This distinctive feature of the CPA allows the minimum output intensity to serve as a highly sensitive indicator for high-performance sensing. In the following, we experimentally verify the enhanced SNR at the CPA EP3 and demonstrate this sensing mechanism in a passive cavity magnonic system.

\vspace{1em}
\noindent\textbf{Enhanced response of frequency and minimum output intensity at the CPA EP3}\\
To realize the CPA EP3, we precisely tuned the magnon-photon coupling by translating the YIG spheres along the $x$-direction with step motors,  and the detunings by rotating their supporting rods ~\cite{ziqi2022} (see Supplementary Materials). With the magnon dissipation rate set to $\gamma/2\pi=3.0$~MHz, the CPA EP3 is identified at $g_{\rm EP3}/2\pi=3.46$~MHz and $\delta_{\rm EP3}/2\pi=1.73$~MHz. A small variation in the external magnetic field is used as perturbations in the experiment, inducing a frequency shift $\Delta_B$ of the Kittel mode in each YIG sphere. The measured output spectra in response to $\Delta_B$ at and away from the CPA EP3 are presented in Figs.~\ref{fig:2}\textbf{e} and \ref{fig:2}\textbf{f}, respectively. The darkest regions correspond to the system at (or close to) CPA under zero or small $\Delta_B$. For $g=0.9g_{\rm EP3}$, the absorption Hamiltonian exhibits non-degenerate eigenvalues, none corresponding to CPA, and the valley frequency only shows a linear shift with $\Delta_B$ (Fig.~\ref{fig:2}\textbf{f}). By contrast, at the EP3 ($g=g_{\rm EP3}$) all three degenerate eigenvalues shift simultaneously, producing a substantially amplified frequency response (Fig.~\ref{fig:2}\textbf{e}). 

\begin{figure*}[t]
	\centering
	\includegraphics[width=0.999\textwidth]{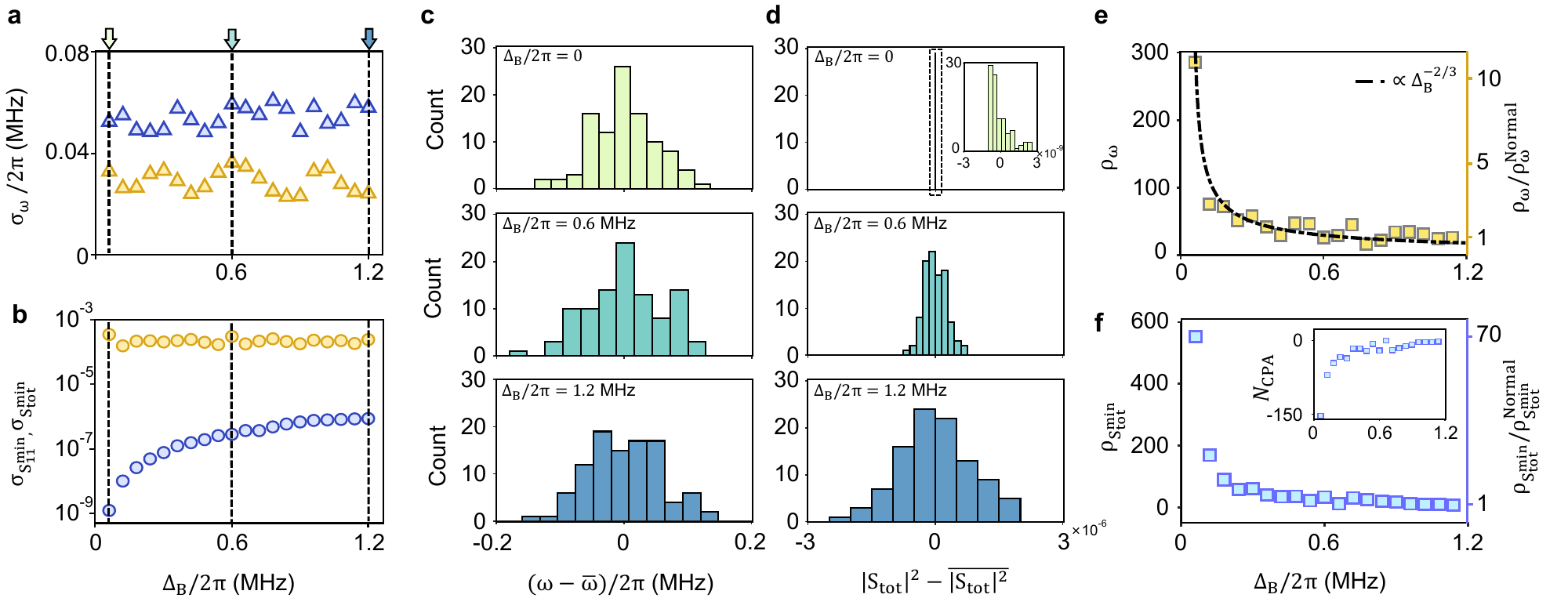}
	\caption{\textbf{Enhanced signal-to-noise ratio of the third-order exceptional point of coherent perfect absorption sensor.} \textbf{a,b}, Noise quantified as the one-standard-deviation uncertainty in frequency $\sigma_\omega$ (\textbf{a}) and in the minimum output intensity $\sigma_{S_{\rm tot}^{\rm min}}, \sigma_{S_{11}^{\rm min}}$ (\textbf{b}) versus the perturbation $\Delta_B$, obtained from one hundred independent repeated measurements. Yellow and blue data correspond to single- and two-port excitations at $g=g_{\rm EP3}$, respectively. \textbf{c,d}, Histograms of the frequency deviations (\textbf{c}) and minimum output intensity deviations (\textbf{d}) at the $\Delta_B$ values indicated by the arrows in \textbf{a} and \textbf{b}, respectively. \textbf{e,f}, Signal-to-noise ratio of the third-order exceptional point of coherent perfect absorption sensor for frequency $\rho_{\omega}$ (\textbf{e}) and minimum output intensity $\rho_{S_{\rm tot}^{\rm min}}$ (\textbf{f}) plotted versus the perturbation $\Delta_B$. The left and right y-axes show the absolute and normalized values, respectively. The inset in \textbf{f} shows the noise reduction factor $N_{\rm CPA}$ calculated from the noise in the minimum output intensity $\sigma_{S_{\rm tot}^{\rm min}}$.} \label{fig:4}
\end{figure*}

By setting the initial valley frequency at the CPA EP3 as the reference ($\Omega^{\rm abs}=0$), a perturbation $\Delta_B$ shifts the valley frequency by $\Delta_\omega \equiv \Omega^{\rm abs'}$, where  $\Omega^{\rm abs'}$ is the eigenvalue of perturbated absorption Hamiltonian. Since only $\Omega_{+}^{\rm abs'}$ lies near the real axis in Fig.~\ref{fig:2}\textbf{a}, this frequency shift can be well approximated as $\Delta_\omega \approx \Omega_{+}^{\rm abs'}$. The experimentally measured $\Delta_\omega$ as functions of $\Delta_B$ at the CPA EP3 (brown) and away from it (blue) are shown in Fig.~\ref{fig:3}\textbf{a}. The extracted data agree well with the theoretical predictions, showing $\Delta_\omega = g^{2/3}\Delta_B^{1/3}$ at the EP3, whereas $\Delta_\omega \sim \Delta_B$ away from it. To directly quantify this enhancement, we characterize the frequency response at the EP3 (Fig.~\ref{fig:3}\textbf{c}) as $F_{\rm EP3}\equiv\left(\frac{\partial\Delta_\omega}{\partial\Delta_B}\right)_{\rm EP3} \propto g^{2/3}\Delta_B^{-2/3}$ when $\Delta_B$ is sufficiently small. At the EP3, $\Delta_\omega$ exhibits a {\it steep} slope compared with the case away from the EP3, leading to a more pronounced amplification of the eigenfrequency shift for the same perturbation $\Delta_B$. For finite perturbations, our measurements demonstrate a fifteen-fold enhancement in the frequency response (brown arrow in Fig.~\ref{fig:3}\textbf{c}).

A remarkably sharp dip expectedly appears in the decibel output spectrum under CPA, but it rapidly flattens once the system slightly deviates from CPA (Fig.~\ref{fig:2}\textbf{e}). This phenomenon is indeed absent in the non-CPA system (Fig.~\ref{fig:2}\textbf{f}). In the normal non-CPA system, $|S_{\rm tot}^{({\rm min})}|^2$ remains nearly unchanged for small $\Delta_B$~(Fig.~\ref{fig:2}\textbf{d}). By contrast, at the CPA EP3, where $|S_{\rm tot}^{({\rm min})}(0)|^2_{\rm CPA}=0$, the $10\log_{10}|S_{\rm tot}^{({\rm min})}(0)|^2_{\rm CPA}$ ideally tends to negative infinity (see Supplementary Materials), giving rise to a divergence of $I_{\rm CPA}$. In practice, the minimum measurable value of $10\log_{10}|S_{\rm tot}^{({\rm min})}(0)|^2_{\rm CPA}$ at CPA EP3 is limited to $-89.8$~dB by the finite resolution of the VNA and imperfections in system alignment. Nevertheless, this minimum still leads to a substantial increase in $\Delta|S_{\rm tot}^{({\rm min})}|^2$ at CPA compared to the non-CPA case, particularly for small $\Delta_B$ (Fig.~\ref{fig:3}\textbf{b}). Although such limitations constrain the divergence rate of $I_{\rm CPA}$, the measured $I_{\rm CPA}$ is still enhanced by two orders of magnitude (blue arrow in Fig.~\ref{fig:3}\textbf{d}).

\vspace{1em}
\noindent\textbf{Enhanced SNR at the CPA EP3 in magnetic field sensing}\\
Although the CPA EP3 enhances the responses in both frequency and minimum output intensity, realizing a high-performance sensor further requires superior measurement precision. This precision is ultimately limited by the sensor’s noise, which sets the bound on the smallest detectable perturbation. To comprehensively evaluate the impact of noise on these observables, we determine the one-standard-deviation (1$\sigma$) uncertainties of the frequency $\sigma_\omega$ and the minimum output $\sigma_{S_{\rm tot}^{\rm min}}$ under different bias magnetic fields. Note that the noise associated with identifying the minimum output location is also included in this analysis, since $|S_{\rm tot}^{({\rm min})}|^2$ is extracted directly rather than obtained through manual spectral fitting.

To examine whether the frequency noise is governed by the completeness of the system eigenbasis, we inject only one probe signal into port~1 at $g=g_{\rm EP3}$. The measured minimum reflection remains nearly constant with perturbation $\Delta_B$, and the corresponding frequency shift follows the eigenvalues of $H_{\rm res}$ (see Supplementary Materials). The standard deviations $\sigma_\omega$ and $\sigma_{S_{11}^{\rm min}}$ obtained from one hundred repeated measurements at various $\Delta_B$ are plotted as yellow points in Figs.~\ref{fig:4}\textbf{a} and \ref{fig:4}\textbf{b}. Both quantities remain nearly unchanged with $\Delta_B$, consistent with the non-divergent PF in this perturbation range.

Still at $g=g_{\rm EP3}$, two configured input signals lead to CPA EP3 while preserving the same eigenbasis as in the single-signal probing. This is evidenced by the nearly invariant $\sigma_\omega$ around the CPA EP3, as shown by the blue points in Fig.~\ref{fig:4}\textbf{a}. To visualize this result more intuitively, Fig.~\ref{fig:4}\textbf{c} presents frequency histograms from repeated measurements at $\Delta_B/2\pi=0$, $0.6$, and $1.2$~MHz (top to bottom), each centered at its mean value $\bar{\omega}$. With identical axes and bin widths, these histograms exhibit comparable distributions and fluctuations, without the abnormally amplified deviations reported in earlier EP-sensor studies~\cite{kononchuk_exceptional-point-based_2022,peters2022exceptional,xia_2025,Lai2019}. Compared with the single-signal case, the slightly higher standard noise near the CPA EP3 is attributed to the inclusion of additional circuit components in the measurement.

For the minimum output noise $\sigma_{S_{\rm tot}^{\rm min}}$, we observe an unexpected and pronounced decrease in the vicinity of CPA, where it is nearly five orders of magnitude lower than in the single-port case, as indicated by the blue points in Fig.~\ref{fig:4}\textbf{b}. The histograms of $|S_{\rm tot}^{({\rm min})}|^2$ are shown in Fig.~\ref{fig:4}\textbf{d}, each centered at its mean value $\bar{|S_{\rm tot}^{({\rm min})}|^2}$, for $\Delta_B/2\pi = 0$, $0.6$, and $1.2$~MHz (top to bottom). Evidently, the fluctuation of $|S_{\rm tot}^{({\rm min})}|^2$ increases appreciably as $\Delta_B$ grows. The inset of Fig.~\ref{fig:4}\textbf{d} further shows that the distribution at $\Delta_B/2\pi=0$ is non-uniform, which arises from the lower detection limit of the VNA, preventing the measurement of smaller values. To identify the dominant noise sources, we measured and fitted the noise as a function of $|S_{\rm tot}|^2$ within a single spectrum at $\Delta_B/2\pi=0$ and $\pm0.06$~MHz. The result shows $\sigma_{S_{\rm tot}} \propto |S_{\rm tot}|$, indicating that shot noise is the main contributor to output intensity noise (see Supplementary Materials), which in turn accounts for the reduction of $\sigma_{S_{\rm tot}^{\rm min}}$ near the CPA EP3.

The SNR for the frequency is defined as $\rho_{\omega}\equiv \bar{F}_{\rm EP3}/\sigma_\omega$, as shown in Fig.~\ref{fig:4}\textbf{e}. In the absence of additional noise, the increase of $\rho_{\omega}$ near the CPA EP3 primarily arises from the growth of $F_{\rm EP3}$, as captured by the fitted curve in Fig.~\ref{fig:4}\textbf{e}. This improvement can be quantified by the normalized SNR. Using the SNR far from the CPA EP3 ($\Delta_B/2\pi=1.2$~MHz) as a reference, the yellow axis on the right side of Fig.~\ref{fig:4}\textbf{e} indicates a {\it twelve-fold} enhancement in SNR near the CPA EP3.

To characterize the variation of noise in the minimum output intensity on a decibel scale, we define the noise reduction factor as $N_{\rm CPA} \equiv \partial\Delta\sigma_{S_{\rm tot}^{\rm min}}/\partial\Delta_B$, with $\Delta\sigma_{S_{\rm tot}^{\rm min}} \equiv 10\log_{10}\sigma_{S_{\rm tot}^{\rm min}}(\Omega^{\rm abs}) - 10\log_{10}\sigma_{S_{\rm tot}^{\rm min}}(\Omega^{\rm abs'})$. The SNR for the minimum output intensity is then defined in decibels as $\rho_{S_{\rm tot}^{\rm min}}\equiv \bar{I}_{\rm CPA}-N_{\rm CPA}$ (see Supplementary Materials). These two factors are shown as functions of the perturbation in the inset and the main panel of Fig.~\ref{fig:4}\textbf{f}, respectively. Due to the combined effect of a substantial increase in $I_{\rm CPA}$ and a marked reduction in $N_{\rm CPA}$, the improvement in $\rho_{S_{\rm tot}^{\rm min}}$ at the CPA EP3 is even more pronounced. Remarkably, the normalized $\rho_{S_{\rm tot}^{\rm min}}$ exhibits a {\it seventy-fold} enhancement near the CPA EP3.

It should be emphasized that the enhancement of $\rho_{\omega}$ at the CPA EP3 is fundamentally limited in our measurements, primarily because the minimum attainable $\Delta_B$ is experimentally constrained. In principle, however, CPA EP3 circumvents the frequency-noise divergence while retaining the nonlinear response of the EP3, thereby enabling a superior SNR compared with conventional EP sensors.

\vspace{1em}
\noindent\textbf{Discussion}\\
In summary, we have realized a noise-resistant higher-order exceptional point (EP) sensor at the coherent perfect absorption (CPA) EP3 in a passive cavity magnonic system, achieving a {\it twelve-fold} enhancement in frequency signal-to-noise ratio (SNR) for magnetic field sensing. Additionally, we exploit the pronounced variation in minimum output intensity near the CPA to enhance sensing and realize a {\it seventy-fold} improvement in SNR. The CPA EP3 is implemented via a pseudo-Hermitian absorption Hamiltonian constructed with two identical YIG spheres in a cavity. For small magnetic field as perturbations, we observe fifteen-fold and 400-fold enhancements in responsivity for frequency and minimum output intensity, respectively. By engineering the system’s absorption EP apart from its resonance EP, we successfully avoid the additional noise arising from the incompleteness of the eigenbasis. A comprehensive noise analysis is conducted as the one-standard deviation of one hundred repeated measurements. Unlike previous EP sensors, our system does not exhibit divergent frequency noise near the CPA EP3. Furthermore, the spectral output intensity noise is notably suppressed near the CPA since it is governed by the shot noise and propotional to the spectra output intensity.

These results establish a robust and tunable platform for EP sensing, opening promising avenues for the development of ultrasensitive, noise-resilient sensors based on both CPA and EP in passive systems. The core idea of separating the two types of EPs in our scheme can be extended to a broad range of experimental platforms. Examples include tailoring the evanescent coupling between a probe fiber and a whispering-gallery-mode microcavity in the optical domain~\cite{Peng2014,Chen2017}, as well as engineering the capacitive coupling between a feedline and an RLC resonator in electronic circuits~\cite{xiao_enhanced_2019,zhao_2024}.
Our approach paves the way for quantum metrology and next-generation sensing technologies~\cite{Degen_2017}.\\

\noindent\textbf{Data Availability}\\
The raw data used in this study are available on Zenodo database [https://doi.org/10.5281/zenodo.18410900].

\providecommand{\noopsort}[1]{}\providecommand{\singleletter}[1]{#1}%

\noindent\textbf{Acknowledgements}\\
This work is supported by the National Natural Science Foundation of China (No.~$\rm{U}25\rm{A}20199$, No.~$92265202$, No.~$12174329$, and No.~$123\rm{B}2064$), the National Key Research and Development Program of China (No.~2022YFA1405200 and No.~2023YFA1406703),  the Zhejiang Provincial Natural Science Foundation of China (No.~$\rm{LR}26\rm{A}040001$) and the ``Pioneer" and ``Leading Goose" R\&D Program of Zhejiang Province (No.~$2025\rm{C}01028$).\\

\noindent\textbf{Author Contributions}\\
J.Q.Y. initiated and supervised the research project. Z.Q.W., Y.D.H. and Y.P.W. designed the experiment. Y.M.S., Z.Q.W., Y.D.H., R.C.S. and W.J.W performed the experimental measurements. Z.Q.W., Y.M.S., Y.D.H., Y.P.W and J.Q.Y analyzed the data, developed the theory, and drafted the manuscript. All authors were involved in the discussion of results and the final manuscript editing.
\\

\noindent\textbf{Competing Interest}\\
The authors declare no competing interest.
\\

\end{document}